\documentclass[prl,reprint, twocolumn,showpacs,preprintnumbers,amsmath,amssymb,nofootinbib,floatfix]{revtex4-1} 

\usepackage{graphicx}
\usepackage{mathrsfs}
\usepackage{hyperref}
\usepackage{slashed}
\usepackage{color}

\newcommand{\beq}{\begin{eqnarray}}
\newcommand{\eeq}{\end{eqnarray}}

\begin{document}

\title{QCD analysis of $B \to \pi\pi$ form factors and the $\vert V_{ub} \vert$ extraction from $B_{l4}$ decays}

\author{Shan Cheng}\email{Corresponding author: scheng@hnu.edu.cn}
\affiliation{School of Physics and Electronics, Hunan University, 410082 Changsha, China}
\affiliation{School for Theoretical Physics, Hunan University, 410082, Changsha, China}
\affiliation{Hunan Provincial Key Laboratory of High-Energy Scale Physics and Applications, 410082 Changsha, China}

\date{\today}

\begin{abstract}

The $B_{l4}$ decays offer an independent method for determining the Cabibbo-Kobayashi-Maskawa matrix element $\vert V_{ub} \vert$, 
with a QCD analysis of the underlying $B \to \pi\pi$ form factors serving as a crucial component.
We present a comprehensive QCD analysis of the $B \to \pi\pi$ form factors within the framework of light-cone sum rules, 
employing two-pion light-cone distribution amplitudes. 
For the first time, we derive the twist-three $2\pi$DAs explicitly by exploiting the conformal symmetry of massless QCD. 
This breakthrough enables the first systematic calculation of higher-power corrections to the $B \to \pi\pi$-type transitions.
Leveraging the recently measured partial decay fractions of $B^+ \to \pi^+\pi^- l^+ \nu_l$ by the Belle detector, 
we obtain $\vert V_{ub} \vert = \left( 4.27 \pm 0.49\vert_{\rm data} \pm 0.55\vert_{\rm LCSRs} \right) \times 10^{-3}$ 
within the dominant region of the $\rho$ resonance. 
We clarify that the relatively small value of $\vert V_{ub} \vert$ extracted from $B \to \rho l \nu$, 
when compared to the golden channel $B \to \pi l\nu$, arises due to the finite width and nonresonant effects. 
We anticipate future measurements of purely isovector and isoscalar $B_{l4}$ decays, 
which will aid in reducing uncertainties by enhancing our understanding of the $\pi\pi$ phase shifts 
in the two-pion distribution amplitudes through heavy flavor decays.
  
\end{abstract}


\maketitle

\textbf{\textit{Introduction.}}--The determination of the Cabibbo-Kobayashi-Maskawa (CKM) matrix element $V_{ub}$ 
is a high priority in the era of precision flavor physics for uncovering new interactions beyond the Standard Model. 
This element plays a crucial role as the dominant source of weak phase difference, 
explaining the significant charge-parity violation observed in hadronic $B$ and $\Lambda_b$ decays \cite{BaBar:2001pki,Belle:2001qdd,LHCb:2025ray}. 
Additionally, its magnitude reflects the transition probability between the mass eigenstates of the third and first quark generations, 
making it the smallest and least precisely known CKM matrix element.
Semileptonic decays induced by the charged transition $b \to u l \nu$ 
offer two avenues to determine $\vert V_{ub} \vert$, namely through the processes with inclusive and exclusive final states. 
For a long time, the inclusive determination has exhibited approximately $2.5 \sigma$ discrepancy  
compared to the exclusive determination \cite{PDG-Vub,HFLAV2023}. 
This inconsistency, known as the $\vert V_{ub} \vert$ puzzle, remains unresolved to date.
 
Compared to the inclusive approach, the exclusive determination of $\vert V_{ub} \vert$ has significantly smaller uncertainty 
due to the superior signal-to-background ratio and the increasingly precise QCD predictions 
\cite{FermilabLattice:2015mwy,Flynn:2015mha,Colquhoun:2022atw,Ball:2004ye,Leljak:2021vte,Cui:2022zwm}. 
Besides the golden channel $B \to \pi l \nu_l$, the $B \to \rho l \nu_l$ and $B_s \to K l \nu_l$ decays, driven by the same $b \to u l \nu$ transition, 
serve as additional compelling channels for the exclusive determination. 
Recently, the Belle II collaboration conducted simultaneous measurements of the partial branching fractions for 
$B^0 \to \pi^- l^+ \nu_l$ and $B^+ \to \rho^0 l^+ \nu_l$ decays \cite{Belle-II:2024xwh}. 
They derived $\vert V_{ub} \vert$ from these two modes by leveraging constraints on the form factors (FFs) from lattice QCD (LQCD) \cite{HFLAV2023} and the light-cone sum rules (LCSRs) \cite{Bharucha:2015bzk}, 
\beq &&\vert V_{ub} \vert_{B \to \pi l \nu} = \left( 3.73 \pm 0.10 \pm 0.16_{\rm LQCD+LCSRs} \right) \times 10^{-3}, \nonumber\\
&&\vert V_{ub} \vert_{B \to \rho l \nu} = \left( 3.19 \pm 0.22 \pm 0.26_{\rm LCSRs} \right) \times 10^{-3}. \label{eq:Vub-BelleII} \nonumber \eeq 
The result derived from the $B \to \rho$ transition aligns with prior analyses \cite{Gao:2019lta,Bernlochner:2021rel,BaBar:2010efp,Belle:2013hlo}, 
yet it is notably small than that determined from the $B \to \pi l \nu_l$ decay. 
In contrast, the $\vert V_{ub} \vert$ extracted from the $B_s \to K^-\mu^+\nu$ decay is largely consistent with the 
golden channel \cite{Gonzalez-Solis:2021pyh,LHCb:2020ist}. 
This inconsistency has raised concerns regarding the uniformity of $\vert V_{ub} \vert$ determinations across different exclusive decays. 

The signal state in a $B \to \rho l \nu_l$-type decay is $\pi\pi l \nu_l$, 
as the $\rho$ is an unstable resonance that decays almost exclusively into $\pi\pi$ state. 
Consequently, a comprehensive analysis must account for the finite widths, 
the nonresonant backgrounds, and the contributions of different partial waves in $\pi\pi$ invariant mass spectrum, 
rather than treating the $\rho$ meson as a single particle at its pole mass \cite{Bharucha:2015bzk} 
or approximating it with a simple Breit-Wigner lineshape \cite{Belle-II:2024xwh}.
Recently, the $B^+ \to \pi^+\pi^- l^+ \nu_l$ decay was measured by the Belle detector \cite{Belle:2020xgu}. 
The branching fraction exhibits a significant deviation from the earlier analysis of $B^+ \to \rho^0 l^+ \nu_l$ decay 
by using the same data set \cite{Belle:2013hlo}, revealing the importance of nonresonant contributions and $S$-wave effects 
which were previously underestimated.

From a QCD perspective, the dynamics of four-body semileptonic decay ($B_{l4}$) \cite{Faller:2013dwa,Kang:2013jaa} 
is governed by the $B \to \pi\pi$ FFs. They were pioneer studied from the light-cone sum rules (LCSRs) approaches. 
Compared to the $B$-meson LCSRs \cite{Cheng:2017smj}, the LCSRs with two-pion light-cone distribution amplitudes ($2\pi$DAs) \cite{Hambrock:2015aor,Cheng:2017sfk,Cheng:2019hpq,Cheng:2019tgh} offer several unique advantages. 
First, the sum rule analysis of the correlation function is performed without resonant modeling assumptions. 
Second, it provides a unified framework to predict contributions from different partial waves.
The last but not the least, theoretical uncertainties are better controlled due to highly suppressed power corrections ${\cal O}(1/m_b)$ \cite{Boer:2016iez}. However, existing LCSRs' studies with $2\pi$DAs have so far been limited to leading-twist accuracy 
due to the incomplete understanding of distribution inside $\pi\pi$ system. This limitation has hindered theoretical progress in $B_{l4}$ decays. 
            
In this work, we derive the subleading-twist $2\pi$DAs 
by rigorously implementing conformal and rotational symmetries in massless QCD. 
These findings represent the first comprehensive advance beyond the two-decades-old leading-twist paradigm \cite{Diehl:1998dk}, providing unprecedented insight into the structure of the $\pi\pi$ system and complementing recent studies of kinematical higher-twist generalized distribution amplitudes (GDAs) \cite{Lorce:2022tiq,Pire:2023ztb,Bhattacharya:2025awq}. 
Our result enable the calculation of subleading power corrections to $B \to \pi\pi$ transition. 
As a key application, we determine $\vert V_{ub} \vert$ from the $B^+ \to \pi^+\pi^- l^+ \nu_l$ channel. 

\textbf{\textit{Two-pion light-cone distribution amplitudes.}}--The most general nonperturbative quantity 
in hard exclusive processes with two energetic pions in the final state is provided by the $2\pi$DAs. 
These amplitudes offer a systematic framework for disentangling resonant (from different partial-waves) 
contributions from nonresonant backgrounds in multi-body heavy-meson decays. 
The $2\pi$DAs are defined by the vacuum to $\pi\pi$ matrix elements sandwiched with special nonlocal quark currents 
\beq \langle \pi^+(k_1) \pi^-(k_2)\vert {\bar q}(0) \Gamma q(x) \vert 0 \rangle 
\propto \int du e^{i {\bar u }k \cdot x} \Phi^{t}({\bar u}, \zeta, k^2). \label{eq:2piDAs} \eeq 
Here $\Gamma \in \{ \gamma_\mu, \gamma_\mu \gamma_5, \sigma_{\mu\nu}, 1 \}$, $k = k_1+k_2$ is the invariant mass of $\pi\pi$ system. 
The superscript $t$ denotes the twist that defined by the difference between canonical dimension 
and conformal spin of the associated quark currents \cite{Balitsky:1987bk,Braun:2003rp}. 
In addition to the longitudinal momentum fraction $u$ carried by the current antiquark, 
hence ${\bar u} = 1-u$ by the quark with respect to the total two-pion momentum, $2\pi$DAs depend on two further kinematic variables: 
the one-pion momentum fraction along the light-cone direction $\zeta = k_1^+/k^+$ , and the invariant mass squared $k^2$. 

Further details on the matrix element definition and twist clarification for $2\pi$DAs are provided in the Supplemental Material. 
Here, we focus on the double expansion formalism. At the parton level, the momentum distributions, charactered by the variable $u$, 
are solved via a local collinear conformal expansion and expressed in terms of Gegenbauer polynomials $C_n^{3/2}(2u-1)$. 
At the hadron level, the kinematic distributions encode the angle momentum of the produced pion pair.
They are decomposed into partial waves and expanded in another set of Gegenbauer polynomials, $C_l^{1/2}(2 \zeta -1)$, 
which is identical to the Legendre polynomials. In practice, the variable $\zeta$ is replaced by the emission angle of a pion in the $\pi\pi$ rest frame 
through the relation $2\zeta -1 = \cos\theta_\pi \beta_\pi(s)$. Here, $\beta_\pi(s) = \left( 1 - 4m_\pi^2/s \right)^{1/2}$ is the phase space factor. 
Consequently, $2\pi$DAs are expressed in a unified form \cite{Diehl:2000uv,Polyakov:1998ze,Lehmann-Dronke:1999vvq} 
\beq &&\Phi \propto \sum_{n}^\infty \sum_{l}^{n+1} B_{nl}(\mu, k^2) C_n^{3/2}(2u-1) C_l^{1/2}(2\zeta-1) \label{eq:2pidas-t2} \eeq 
with some certain normalization constraints at different twists. 
Here $n$ is even (odd) and $l$ is odd (even) for the isovector (isoscalar) $\pi\pi$ state. 

The expansion coefficients $B_{nl}(\mu, k^2)$ 
weights the universal distribution amplitudes at a given conformal power $n$ and partial wave $l$. 
They encode both renormalization scale dependence and invariant mass dynamics via 
\beq B^{\parallel}_{nl}(\mu, k^{2}) &=& B^{\parallel}_{nl}(0) \left[ \frac{\alpha_s(\mu)}{\alpha_s(\mu_0)}\right]^{\frac{\gamma_n - \gamma_0}{2\beta_0}} 
{\rm Exp} \Big[ k^{2} \frac{d \ln B^{\parallel}_{nl}(0)}{dk^{2}} \nonumber\\
&~&  + \frac{k^{4}}{\pi} \int_{4m_\pi^2}^\infty ds \frac{\delta_l(s)}{s^2 (s-k^{2}-i0)} \Big]. \label{eq:Bnl} \eeq 
The scale dependence of the coefficients, governed by the anomalous dimension $\gamma_{n}$ that identifies to the single-meson case \cite{Chernyak:1977as,Lepage:1979zb,Efremov:1979qk}, 
reflects shrinking transverse parton separations at high momentum \cite{Ball:1996tb,Ball:1998sk,Ball:1998ff,Ball:2007zt}. 

Near the threshold, the parameters $B_{nl}(0)$ and $d \ln B^{\parallel}_{nl}(0)/dk^{2}$ can be determined 
from the instanton model of the QCD vacuum. 
The instanton vacuum is modeled as a grand canonical ensemble, 
with the partition function expressed in terms of the collective coordinates.  
Within this framework, the mean values of these coordinates are determined 
by maximizing the partition function for a fixed number of pseudoparticles. 
A key element of this derivation is the introducing of a dimensionless parameter to incorporates the argument of the inverse charge 
that is not uniquely fixed from first principles. As indicated in Table I of Ref. \cite{Diakonov:1995qy}, 
the resulting uncertainties in the average instanton size $\bar{\rho}$ and the average instanton separation $\bar{R}$ are each approximately $10\%$. However, their ratio $\bar{\rho}/\bar{R} \sim 1/3$ is notably stable against variations in this parameter. 
In our analysis, we adopt the established values from Refs. \cite{Shuryak:1982dp,Petrov:1998kg}, 
allowing the associated uncertainties to be safely neglected in the construction of the $2\pi$ distribution amplitudes. 

For the resonant regions, Watson's theorem yields the $k^2$-dependence via Omn\'es solutions with 
the phase shifts $\delta_l$ in pion-pion scattering \cite{Omnes:1958hv}, implemented through subtracted dispersion relations. 
In practice, an excellent description of data up to $k^2 \sim 5$ GeV$^2$ is achieved through 
an amplitude analysis that combines dispersion relations with unitarity constraints \cite{Dai:2014zta}. 
This analysis fits the phase shifts and inelasticity parameters to $\pi^- p \to \pi^-\pi^+ n$ hadronic data from threshold up to approximately $1.8$ GeV. 
Final-state interactions between meson pairs, especially critical in the $K{\bar K}$ threshold region, 
are further constrained by BABAR Dalitz plot analyses of $D_s^+ \to \pi^+ \pi^- \pi^+$ and $K^+K^- \pi^+$ \cite{Grayer:1974cr,BaBar:2008nlp,BaBar:2010wqe}.  
At higher energies, the full $\gamma\gamma \to \pi\pi$ amplitudes are described using a Regge parameterization \cite{Pelaez:2004vs,Kaminski:2006yv,Garcia-Martin:2011iqs}. 

In exclusive QCD processes, power-suppressed contributions arise from three sources: 
the "bad" component of the quark field with an incorrect spin projector, 
transverse momentum in the leading-twist components of the $q{\bar q}$configuration, 
and higher Fock states such as $q{\bar q}g$ and $q{\bar q} q {\bar q}$. 
These contributions are encoded in subleading-twist LCDAs. 
For $2\pi$DAs, the quark current operators match those of single-meson LCDAs (with the same quantum numbers), 
leading to three analogous terms in the two-particle twist-three $2\pi$DAs: 
a term connected to three-particle $2\pi$DAs through equation of motions (EOM), 
a quark mass-dependent term, and a part related to the leading twist DAs. 
In this work, we focus exclusively on the latter contribution, referred to as the Wandzura-Wilczek (WW) contribution, for twist-three $2\pi$DA. 
The remaining two terms are neglected since the three-particle $2\pi$DAs, though unexplored, 
are expected to be power suppressed as ${\cal O}(1/m_b)$ in $B$ decays, while the light-quark mass effects are negligible at ${\cal O}(m_q/m_\pi)$. 
Our definition of twist-three $2\pi$DAs preserves the scale dependence and normalization of the single-meson parton momentum distribution, 
augmented by partial-wave polynomials at Gegenbauer order $n$ with coupling coefficient $B_{nl}(\mu, k^2)$. 
Explicit expressions of two-particle twist-three $2\pi$DAs of both isovector and isoscalar $\pi\pi$ systems are provided in the Supplemental Material.

\textbf{\textit{$B \to \pi\pi$ form factors.}}--The transition FFs between the $B$ meson and the $\pi\pi$ system 
are defined through the matrix elements of the weak current, 
\beq &~&\langle \pi^+(k_1) \pi^-(k_2) \vert {\bar u} \gamma^\mu \left( 1 - \gamma_5 \right) b \vert {\bar B}^0(p_B) \rangle \nonumber\\
&=& i F_\perp \frac{{\bar q}_\perp^\mu}{\sqrt{k^2}} + F_t \frac{q^\mu}{\sqrt{q^2}} 
+ F_0 \frac{2 \sqrt{q^2}  k_0^\mu}{\sqrt{\lambda}} + F_\parallel \frac{{\bar k}^\mu_{\parallel}}{\sqrt{k^2}}. \label{eq:B2pipi-ff} \eeq
The orthogonal basis of momentum vectors \cite{Faller:2013dwa} are    
${\bar q}_\perp^\mu = 2 \varepsilon^{\mu\nu\rho\sigma}q_\nu k_\rho {\bar k}_\sigma/\sqrt{\lambda}$, 
$q^\mu = p_B^\mu - k^\mu$, $k_0^\mu = k^\mu - k\cdot q/q^2 q^\mu$ and 
${\bar k}^\mu_{\parallel} = {\bar k}^\mu - 4 (k \cdot q) ({\bar k} \cdot q)/\lambda k^\mu + 4 k^2 ({\bar k} \cdot q)/\lambda q^\mu$.
The FFs $F_{i}$ with $i=\{\perp,\parallel,t,0\}$ depend on three kinematical variables: 
momentum transfer squared $q^2$, dipion invariant mass squared $k^2$, 
and scalar produce $q \cdot {\bar k} = \beta_\pi\cos \theta_\pi \sqrt{\lambda}/2$ 
that specifies the polar angle $\theta_\pi$ of the $\pi^-$ meson in the $\pi\pi$ rest frame. Here ${\bar k} = k_1 - k_2$, 
the K\"all\'en function reads as $\lambda = m_B^4 - q^4 - k^4 - 2 \left( m_B^2 q^2 + m_B^2 k^2 + q^2 k^2 \right)$. 

The derivation of $B \to \pi\pi$ FFs begins with the construction of approximate correlation functions  \cite{Hambrock:2015aor,Cheng:2017sfk,Cheng:2019hpq}
\beq &&C[\Gamma_\mu] = i \int d^4x e^{- i p_B \cdot x} \langle \pi^+ \pi^- \vert T\left\{ j_\mu(0), j_{(B)}(x) \right\} \vert 0 \rangle, \nonumber\\
&&C[\Gamma_5] = i \int d^4x e^{- i p_B \cdot x} \langle \pi^+ \pi^- \vert T\left\{ j_5(0), j_{(B)}(x) \right\} \vert 0 \rangle, \label{eq:cfs} \eeq
in which the weak current $j_\mu=\overline{q}_{f} \gamma_\mu(1-\gamma_5)b$ (pseudoscalar current $j_5 = i m_b\overline{q}_{f}\gamma_5 b$) 
and the $B$-meson interpolating current $j_{(B)} = im_b\overline{b} \gamma_5 q_{f}$, with $f$ being the flavor index, 
are sandwiched between the vacuum and $\pi\pi$ state. 
The QCD calculation of the correlation functions is performed in the negative half-plane of $q^2$, 
which requires the momentum transfer squared to satisfy $0 \leqslant \vert q^2 \vert \leqslant q^2_{\rm max} \sim 10$ GeV$^2$. 
We decompose them into four invariant form factor densities, 
says $C_{\perp}, C_{t}, C_{0}, C_{\parallel}$, associated to the momentum vectors defined in Eq. (\ref{eq:B2pipi-ff}). 
In which, $C_{\perp,\parallel}$ and $C_{t}$ can be directly extracted from the correlation functions $C[\Gamma^\mu]$ and $C[\Gamma_5]$. 
To determine $C_{0}$, we decompose $C[\Gamma^\mu]$ in terms of the momentum vectors 
of the two pions and the weak current (${\bar q}_\perp^\mu, k^\mu, {\bar k}^\mu, q^\mu$), and multiply it with $q_\mu$. 
$C_0$ is ultimately expressed in terms of $C_\perp, C_\parallel$ and an additional density $C_R$. 

On the other hand, as $q^2$ shifts from large negative to positive values, the average distance between the two currents increases, 
leading to the formation of hadrons through long-distance quark-gluon interactions. 
In this context, the correlation functions can be expressed as the sum of contributions from all possible intermediate states. 
$B \to \pi\pi$ FFs are then extracted from the density functions $C_{i}$ by applying the Borel transformation 
and subsequently performing the continuum subtraction. 
\begin{widetext} 
\beq && F_\parallel(q^2,k^2,\theta_\pi) = \frac{- i m_b^2 \sqrt{k^2}}{2 m_B^2 f_B} \int_{u_0}^1 du 
\Big[ \frac{( m_b^2 - q^2 + u^2k^2 ) \Phi_\perp}{ u^2 m_b f_{2\pi}^\perp \beta_\pi \cos\theta_\pi} 
+ \frac{\Phi_\perp^{(v)}}{u} \Big] e^{\frac{m_B^2-s(u,q^2)}{M^2}}, 
\label{eq:Fpara} \nonumber\\
&&F_\perp(q^2,k^2,\theta_\pi) = \frac{i m_b^2 \sqrt{\lambda k^2}}{2 m_B^2 f_B} \Big\{ \int_{u_0}^1 du 
\Big[ \frac{\Phi_\perp}{u m_b f_{2\pi}^\perp \beta_\pi \cos\theta_\pi } 
+ \frac{ \Phi_\perp^{(a)}}{2 u^2 M^2} \Big] e^{\frac{m_B^2-s(u,q^2)}{M^2}} 
+ \frac{\Phi_\perp^{(a)}(u_0) }{2 u_0 ( s_0 -q^2) } e^{\frac{m_B^2-s_0}{M^2}} \Big\}, \label{eq:Fperp} \nonumber \\\
&&\sqrt{q^2} F_t(q^2,k^2,\theta_\pi) = \frac{- i m_b^2}{2 m_B^2 f_B} \Big\{ \int_{u_0}^1 du 
\Big[ \frac{(m_b^2 - q^2 + u^2 k^2) \Phi_\parallel}{u^2} + \frac{ \sqrt{\lambda} \beta_\pi \cos\theta_\pi \Phi_\perp^{(v)} }{u}  \nonumber\\
&& \hspace{1cm} - \frac{m_b k^2 }{f_{2\pi}^\perp} \left( \frac{1}{u^2} - \frac{m_b^2 - q^2 + u^2k^2}{u^3 M^2} \right) \Phi_\parallel^{(s)} \Big] 
e^{\frac{m_B^2-s(u,q^2)}{M^2}} 
+ \frac{m_b k^2 \left(m_b^2 - q^2 + u_0^2k^2 \right) }{f_{2\pi}^\perp u_0 (s_0 - q^2)} \Phi_\parallel^{(s)}(u_0) e^{\frac{m_B^2-s_0}{M^2}}, 
\label{eq:Ft} \nonumber\\
&& \sqrt{q^2} F_0(q^2,k^2,\theta_\pi) = \frac{\sqrt{\lambda} \sqrt{q^2} }{m_B^2 - q^2 -k^2} F_t
+ \frac{2 q^2 \sqrt{k^2} \beta_\pi \cos\theta_\pi}{m_B^2 - q^2 -k^2} F_\parallel 
+ \frac{\sqrt{\lambda} q^2}{i m_B (m_B^2 - q^2 -k^2)} F_R, \label{eq:F0} \nonumber\\
&&F_R(q^2,k^2,\theta_\pi) = \frac{m_b k^2 }{2 f_{2\pi}^\perp m_B f_B} \Big\{ 
\int_{u_0}^1 du \left( \frac{1}{u^2} - \frac{m_b^2 - q^2 + u^2k^2}{u^3 M^2} \right) 
\left[\frac{\Phi_\parallel^{(s)}}{2} + 2 I_1[\Phi_\parallel^{(t)}] \right] e^{\frac{m_B^2-s(u,q^2)}{M^2}} \nonumber \\
&& \hspace{1cm} - \frac{ \left( m_b^2 - q^2 + u_0^2k^2\right)}{u_0 (s_0 - q^2)} 
\left[\frac{\Phi_\parallel^{(s)}(u_0)}{2} + 2 I_1[\Phi_\parallel^{(t)}(u_0)] \right] e^{\frac{m_B^2-s_0}{M^2}} \Big\}. 
\label{eq:FR} \eeq 
\end{widetext} 

Eq. (\ref{eq:FR}) is the main result of this work, where the contributions from twist-three $2\pi$DAs 
($\Phi_\perp^{(a)},\Phi_\perp^{(v)},\Phi_\parallel^{(s)},\Phi_\parallel^{(t)}$) are studied as the first time. 
Our leading twist results ($\Phi_\perp, \Phi_\parallel$ terms) are consistent with the previous works 
\cite{Hambrock:2015aor,Cheng:2017sfk,Cheng:2019hpq}. 
We consider the virtuality of $b$-quark at $s(u,q^2) = (m_b^2 - {\bar u}q^2 + u {\bar u}k^2)/u $, 
and $u_0$ is the solution of $u$ by taking $s = s_0^B$. 
We take the sum rules parameters $s_0^B = 37.5 \pm 2.5$ GeV$^2$ and $M^2 = 16 \pm 4$ GeV$^2$ 
as the same in the previous leading twist study \cite{Hambrock:2015aor,Cheng:2017sfk,Cheng:2019hpq}, 
other input parameters are all the same as in Ref. \cite{Cheng:2019hpq} and the references there. 
Additionally, we introduce an auxiliary function $I_1[\Phi_\parallel^{(t)}]  = \int_0^u du^\prime \Phi_\parallel^{(t)}(u^\prime, \zeta, k^2)$. 

\textbf{\textit{$\vert V_{ub} \vert$ extraction.}}--The triple-differential decay width of $B_{l4}$ process, without considering the lepton mass, is
\beq && \frac{d^3 \Gamma}{dq^2 dk^2 d\cos\theta_\pi } 
= G_F^2 \vert V_{ub} \vert^2\frac{ \beta_\pi \sqrt{\lambda} q^2}{4 \left( 4\pi \right)^5 m_B^3}  \nonumber\\
&& \hspace{1cm} \Big[ \frac{2}{3} \vert F_0 \vert^2 + \frac{2 \beta_\pi^2 \sin^2 \theta_\pi}{3} 
\left(\vert F_\parallel \vert^2 + \vert F_\perp \vert^2 \right)  \Big]. \eeq
The FFs $F_{i}(q^2,k^2,\theta_\pi)$ can be expanded in terms of associated Legendre polynomials $P_l^{(m)}$ as
\beq &&F_{\parallel,\perp}(q^2,k^2,\theta_\pi) = 
\sum_{l=1}^\infty \sqrt{2l+1} F_{\parallel,\perp}^{(l)}(q^2,k^2) \frac{P_l^{(1)}(\cos \theta_\pi)}{\sin\theta_\pi}, \nonumber\\
&&F_{0,t}(q^2,k^2,\theta_\pi) = \sum_{l=0}^\infty \sqrt{2l+1} F_{0,t}^{(l)}(q^2,k^2) P_l^{(0)}(\cos \theta_\pi), \eeq 
with $l = (0,1,\cdots)$ corresponding to $(S,P,\cdots)$-wave contributions. 
By integrating over the polar angle $\theta_\pi$, the double-differential decay width is obtained. 
\beq &&\frac{d^2 \Gamma}{dq^2 dk^2} = G_F^2 \vert V_{ub} \vert^2\frac{\beta_\pi \sqrt{\lambda} q^2}{3 \left( 4\pi \right)^5 m_B^3}  
\Big[ \sum_{l=0,1,\cdots} \vert F_0^{(l)}(q^2,k^2) \vert^2 \nonumber\\
&& \hspace{0.5cm} + \beta_\pi^2 \sum_{l=1,3, \cdots} \left( \vert F_\parallel^{(l)}(q^2,k^2) \vert^2 + \vert F_\perp^{(l)}(q^2,k^2) \vert^2  \right) \Big]. \label{eq:Bl4-2D} \eeq 

The next task is to obtain the $B \to \pi\pi$ FFs for a specific partial wave $F_i^{(l)}(q^2,k^2)$ from the sum rules result given in Eq. (\ref{eq:FR}). 
This is achieved by utilizing the orthogonality properties of the associated Legendre polynomials \cite{Cheng:2017sfk,Cheng:2019hpq}. 
In details, by multiplying $F_{\perp, \parallel}$ by $P_{l'}^{(1)}(\cos\theta_\pi) \sin\theta_\pi$ and $F_t$ by $P_{l'}^{(0)}(\cos\theta_\pi)$, 
and subsequently integrating over $\theta_\pi$. 
We find that the two-particle twist-three contributions vanish in the partial-wave FFs $F_{i=\parallel,\perp}^{(l=1,3,\cdots)}$ 
due to our choice of partial-wave projectors. Specifically, for the transverse-helicity FFs $F_{\parallel,\perp}^{(l=1,3)}$, 
the projector $P_{l'}^{(1)}(\cos\theta_\pi) \sin\theta_\pi$ generates twist-3 integrands 
that are odd functions, causing their integrals over the polar angle to vanish. 
In contrast, these contributions persist for the timelike- and longitudinal-helicity FFs $F_{t,0}^{(l=0,1,\cdots)}$, 
where the projector $P_{l'}^{(0)}(\cos\theta_\pi)$ preserves the even symmetry of the twist-three integrands. 
Notably, the $\Phi_\parallel^{(s)}$ terms provide a significant correction ($\sim 40 \%$) to the leading-twist result in this channel. 

While finite twist-three contributions are present in $B \to \rho$ FFs \cite{Ball:2004rg,Bharucha:2015bzk}, 
we find no such contribution to the $P$-wave transverse-helicity form factors $F_{\parallel,\perp}^{(l=1)}$. 
This difference is attributed to our treatment of the twist-three $2\pi$DAs, 
for which we consider only the WW term, as stated at the end of last section. 
In contrast, the LCSR calculations for $B \to \rho$ transition incorporate a substantial admixture of twist-two matrix elements 
into the twist-three distribution amplitudes (DAs) via the EOM. 
Although this admixture complicates a naive twist-counting of the relative size of contributions of different DAs, 
the genuine twist-three contributions to the $B \to \rho$ FFs, primarily arising from these EOM terms, 
are indeed suppressed by ${\cal O}(m_\rho/m_b)$. 

\begin{figure}[t]
\begin{center} 
\includegraphics[width=0.23\textwidth]{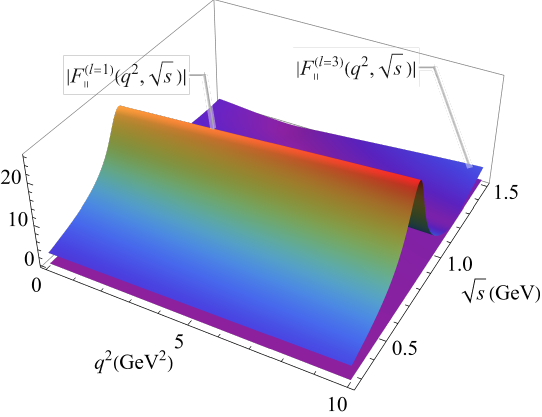}  \hspace{2mm}
\includegraphics[width=0.23\textwidth]{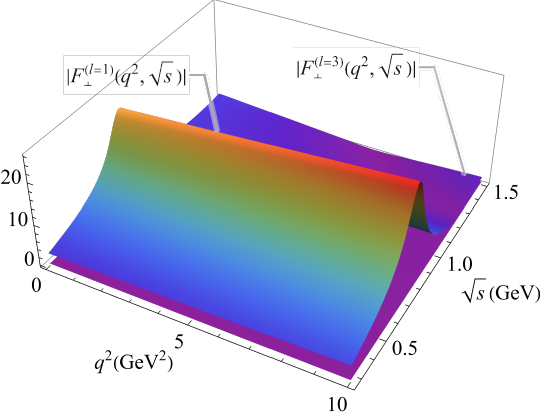} \\ \vspace{4mm}
\includegraphics[width=0.23\textwidth]{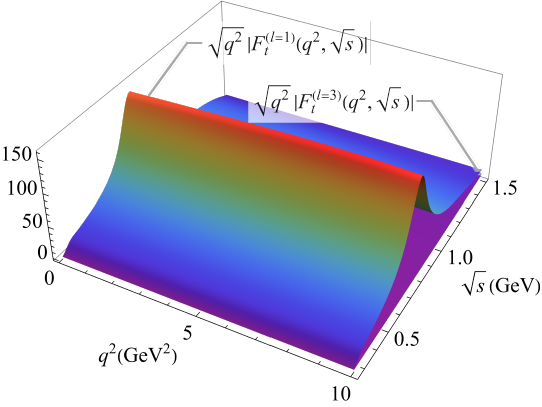}  \hspace{2mm}
\includegraphics[width=0.23\textwidth]{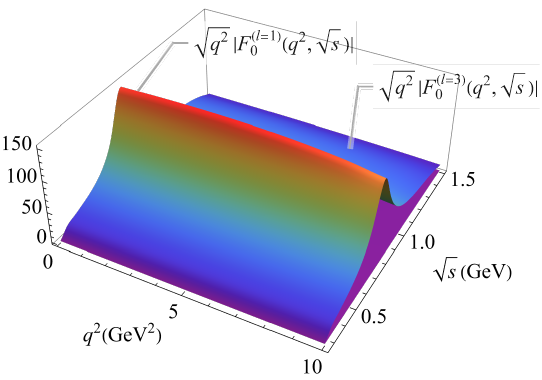} 
\end{center}
\vspace{-4mm}
\caption{$P$- and $F$-wave components to $B \to \pi^+\pi^-$ FFs. }
\label{fig:FFs-P-F}
\end{figure} 
\begin{figure}[t]
\begin{center}  
\includegraphics[width=0.22\textwidth]{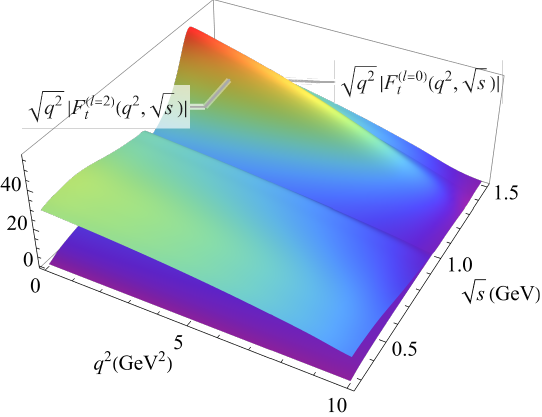}  \hspace{2mm}
\includegraphics[width=0.22\textwidth]{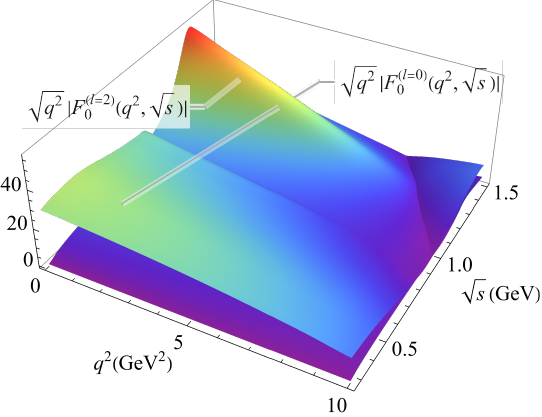} 
\end{center}
\vspace{-4mm}
\caption{$S$- and $D$-wave components of timelike- and longitudinal-helicity FFs. }
\label{fig:FFs-S}
\end{figure} 

In figures \ref{fig:FFs-P-F} and \ref{fig:FFs-S}, we present the $P,F$- and $S,D$-wave components of $B \to \pi^+\pi^-$ FFs, respectively. 
While the $P$-wave contribution dominates in the region of the $\rho$ resonance, 
we observe that the $S$-wave component provides a sizable contribution when the $\pi\pi$ invariant mass is not far away from the threshold value, 
and the $D$-wave component give a significant contribution when the $\pi\pi$ invariant mass is large. 
In contrast, the $F$-wave component is relatively small in the considered kinematic regions.

Using the PDG value of $\vert V_{ub} \vert = \left( 3.82 \pm 0.20 \right) \times 10^{-3}$ \cite{PDG2024}, 
we calculate the partial branching fractions of $B^+ \to \pi^+\pi^- l^+ \nu$. 
The results, presented in table \ref{tab:partial-BF}, are compared with the Belle measurements \cite{Belle:2020xgu}. 
The first and second uncertainties in our calculation arise from $\vert V_{ub} \vert$ and the LCSRs parameters 
($s_0^B$ and $M^2$), respectively. 
We estimate a $\sim 20 \%$ uncertainty in the branching ratios from the LCSR approach, dominated by variations in $M^2$. 
This translates to a $\sim 10\%$ uncertainty in the FFs themselves. 
Given that LCSRs are most reliable at small momentum transfer squared for heavy-to-light transitions, 
we consider seven bins with varying invariant mass ($\sqrt{s} = \sqrt{k^2}$) across different resonant regions, 
while restricting $q^2$ to the range $[0,8]$ GeV$^2$. 

\begin{table}[t]\begin{center} \vspace{-4mm}
\caption{Partial branching fractions $\triangle {\cal B}^i$ (in  unit of $10^{-5}$).} \vspace{2mm}
\label{tab:partial-BF}
\begin{tabular}{l | c c | c c }
\hline\hline
{\rm bins} \quad & $\sqrt{s}$ & $q^2$ & \quad $\triangle {\cal B}^i$ \quad & \quad $\triangle {\cal B}^i$ \cite{Belle:2020xgu} \nonumber\\ \hline
$1$ & $[4m_\pi^2, 0.6]$ & $[0, 8]$ & $0.27 \pm 0.03 \pm 0.06$ & $0.84^{+0.39}_{-0.32}\pm0.18$ \nonumber\\ \hline
$2$ & $(0.6, 0.9]$ & $[0, 4]$ & $1.91 \pm 0.21 \pm 0.38 $ & $2.39^{+0.53}_{-0.47}\pm0.32$ \nonumber\\ 
$3$ & $(0.6, 0.9]$ & $(4, 8]$ & $1.54 \pm 0.17 \pm 0.27 $ & $2.16^{+0.47}_{-0.42}\pm0.23$ \nonumber\\ \hline
$4$ & $(0.9, 1.2]$ & $[0, 4]$ & $0.65 \pm 0.07 \pm 0.12 $ & $0.70^{+0.32}_{-0.25}\pm0.20$ \nonumber\\ 
$5$ & $(0.9, 1.2]$ & $(4, 8]$ & $0.41 \pm 0.04 \pm 0.08$ & $0.64^{+0.28}_{-0.22}\pm0.11$ \nonumber\\ \hline
$6$ & $(1.2, 1.5]$ & $[0, 4]$ & $0.57 \pm 0.05 \pm 0.10$ & $0.91^{+0.35}_{-0.28}\pm0.12$ \nonumber\\ 
$7$ & $(1.2, 1.5]$ & $(4, 8]$ & $0.16 \pm 0.02 \pm 0.02$ & $0.64^{+0.32}_{-0.26}\pm0.08$ \nonumber\\ \hline\hline
\end{tabular}\end{center} \vspace{-4mm} \end{table}

We find good consistent in bins $"2", "3"$ and $"4","5"$, where the invariant mass lies within the $\rho$ and $f_0$ resonances region, respectively. 
In this region, the $2\pi$DAs is well understood, as the (predominant) asymptotic term is proportional to the well measured timelike FFs of the pion \cite{BaBar:2012bdw,Belle:2006wcd}. 
For bin $"1"$, where the invariant mass primarily resides in the region of light scalar resonance $\sigma$, 
our result is approximately one-third of the experimental data. 
This discrepancy can be attributed to two main factors:
(1) The nonasymptotic terms in the twist-three $2\pi$DAs of isoscalar $\pi\pi$ system 
have not yet been studied and are therefore not included in the current calculation, 
but these terms are known to play a significant role in $D \to S$ transitions, where $S$ denotes a scalar meson \cite{Cheng:2019tgh,Cheng:2023knr}. 
(2) The need for a better understanding of the asymptotic term of the isoscalar $2\pi$DAs in the $\sigma$ resonance region, 
which is directly related to the $\pi\pi$ scattering phase shifts. 
In the larger invariant mass region $1.2 \leq \sqrt{s} \leqslant 1.5$ GeV, particularly in bin "8," 
our results are significantly smaller than the experimental measurements. 
This discrepancy underscores the limited knowledge of $2\pi$DAs in this region. 
Furthermore, since the intermediate resonances in this region are not ground states, the predictive power of LCSRs may be reduced. 

The good consistency observed in bins $"2"$ and $"4"$ motivates us to extract $\vert V_{ub} \vert$ 
from the $B_{l4}$ decay in the regions of $\rho$ and $f_0$ resonances. 
We obtain $\vert V_{ub} \vert =  \left( 4.27 \pm 0.49 \pm 0.55 \right) \times 10^{-3}$ and 
$ \left( 3.96 \pm 0.47 \pm 0.52 \right) \times 10^{-3}$ in these $\pi\pi$ invariant mass regions. 
The result are slightly larger but remain consistent with that obtained from the golden channel $B \to \pi l\nu$, 
compared to that obtained from the $B \to \rho l \nu$ channel. 
The first and second uncertainties originate from the experimental data and LCSRs prameters,  
they are notably large at the current precision level.
We anticipate that a complete LQCD simulation at large momentum transfer squared could 
reduce the theoretical uncertainty by approximately half, similar to the improvement seen in the $B \to \pi l\nu$ decay, 
where a combined analysis of LQCD and LCSRs significantly enhanced the precision.  
Recently, the $B \to \pi\pi$ transition FFs in the $\rho$ resonance region have been evaluated using LQCD 
with a simple Breit-Wigner model \cite{Leskovec:2022ubd,Leskovec:2025gsw}. 
A novel parameterization of the $B \to \pi\pi$ form factors, 
based on partial-wave decompositions and series expansions, is proposed for the full phase space \cite{Herren:2025cwv}. 
Consequently, the purely isovector decay $B^0 \to \pi^- \pi^0 l^+ \nu_l$ and the isoscalar decay $B^+ \to \pi^0\pi^0 l^+ \nu_l$ 
are highly anticipated at the Belle II detector. 
These measurements will help distinguish between $P$- and $S$-wave contributions and improve our understanding of $2\pi$DAs.

\textbf{\textit{Conclusions.}}--In this work, we perform the comprehensive calculation of the $B \to \pi\pi$ FFs using LCSRs, 
incorporating an investigation of the subleading twist-three $2\pi$DAs. 
We demonstrate that these previously neglected contributions play a significant role, particularly in the timelike- and longitudinal-helicity FFs, 
highlighting the importance of higher-twist effects in multi-body $B$-meson decays.
Given the well-understood predominance of the $P$-wave contribution in the $B_{l4}$ decay $B^0 \to \pi^+\pi^- l^+ \nu_l$, 
we extract the CKM matrix element $\vert V_{ub} \vert$ via the $B_{l4}$ decays. 
Our result, $\vert V_{ub} \vert = \left( 4.27 \pm 0.49\vert_{\rm data} \pm 0.55\vert_{\rm LCSRs} \right) \times 10^{-3}$,  
is in consistent with that obtained from the golden channel $B \to \pi l \nu_l$. 
This consistency resolves the puzzle regarding the determination of $\vert V_{ub} \vert$ from different exclusive channels, 
particularly in light of finite width and nonresonant effects. 
The uncertainties in our determination can be further reduced through advancements in both theoretical frameworks and experimental precision. 
It is also feasible to perform a next-to-leading-order (NLO) calculation for the $B \to \pi\pi$ form factors
which would significantly enhance the accuracy of the QCD predictions \cite{Ball:1998kk,Duplancic:2008ix,SentitemsuImsong:2014plu}. 
Consequently, $B_{l4}$ decays offer a robust and independent method to address the $\vert V_{ub} \vert$ puzzle, 
providing a critical cross-check for existing determinations.

{\it Acknowledgments:} We are grateful to Vladimir Braun, Wen-bing Qian and Wei Wang for fruitful discussions. 
We also wish to Ling-yun Dai for sharing the original result of pion-pion phase shifts from amplitude analysis \cite{Dai:2014zta}. 
This work is supported by the National Key R$\&$D Program of China under Contracts No. 2023YFA1606000 
and the National Science Foundation of China (NSFC) under Grant No. 12575098 and No. 11975112.

\pagebreak
\widetext
\begin{center} 
\vspace{6mm}
\textbf{\large Supplemental Materials to QCD analysis of $B \to \pi\pi$ form factors and the $\vert V_{ub} \vert$ extraction from $B_{l4}$ decays}
\end{center}
\setcounter{equation}{0}
\setcounter{figure}{0}
\setcounter{table}{0}
\setcounter{page}{1}
\makeatletter
\renewcommand{\theequation}{S\arabic{equation}}
\renewcommand{\thefigure}{S\arabic{figure}}
\renewcommand{\bibnumfmt}[1]{[S#1]}

\appendix

\section{QCD definitions and double expansions of $2\pi$DAs}

The $2\pi$DAs associated to the valence quark state are defined by the nonlocal matrix elements via 
\beq &&\langle \pi^+(k_1)\pi^-(k_2) \vert {\bar q}(0) \gamma^\mu q(x) \vert 0 \rangle = \int du e^{i {\bar u} k \cdot x} 
\Big[ k^\mu \Phi_\parallel({\bar u}, \zeta, k^2) + {\bar k}^\mu \Phi_\perp^{(v)}({\bar u}, \zeta, k^2) \Big], \nonumber\\
&&\langle \pi^+(k_1)\pi^-(k_2) \vert {\bar q}(0) \gamma^\mu \gamma_5 q(x) \vert 0 \rangle = \int du e^{i {\bar u} k \cdot x} 
\frac{\varepsilon^{\mu\nu\rho\sigma}}{4} {\bar k}_\nu k_\rho x_\sigma \Phi_\perp^{(a)}({\bar u}, \eta, k^2), \nonumber\\
&& \langle \pi^+(k_1)\pi^-(k_2) \vert {\bar q}(0) \sigma^{\mu\nu} q(x) \vert 0 \rangle = \int du e^{i {\bar u} k \cdot x} 
\Big[ - \frac{i}{f_{2\pi}^\perp} \frac{k_\mu {\bar k}_\nu - k_\nu {\bar k}_\mu }{2 \zeta -1} \Phi_\perp({\bar u}, \zeta, k^2) 
+ \frac{i k^2}{f_{2\pi}^\perp} \frac{k_\mu x_\nu - k_\nu x_\mu}{k \cdot x} \Phi_\parallel^{(t)}({\bar u}, \zeta, k^2) \Big],\nonumber\\
&& \langle \pi^+(k_1)\pi^-(k_2) \vert {\bar q}(0) q(x) \vert 0 \rangle = \int du e^{i {\bar u} k \cdot x} 
\frac{i k^2 (k \cdot x) }{2 f_{2\pi}^\perp} \Phi_\parallel^{(s)}({\bar u}, \zeta, k^2). \label{eq:2pidas-definition}\eeq
Here ${\bar u} = 1 -u$ is the longitudinal momentum fraction carried by the quark,  
and $f_{2\pi}^\perp$ is the decay constant defined by the local matrix element
$\langle \pi^+(k_1)\pi^-(k_2) \vert {\bar q} \sigma^{\mu\nu} q \vert 0 \rangle 
 \stackrel{k^2 \to 0}{\longrightarrow} -i \left( k_\mu {\bar k}_\nu - k_\nu {\bar k}_\mu \right)/(2 f_{2\pi}^\perp)$. 
$2\pi$DAs are decomposed in eigenfunctions of the evolution equation (Gegenbauer polynomials $C_n^{3/2}(2u-1)$) 
and in partial waves of the produced pions (Legendre polynomials $C_l^{1/2}(2 \zeta-1)$), 
the scale dependence is described by renormalization group methods \cite{Chernyak:1977as,Lepage:1979zb,Efremov:1979qk} 
and absorbed into the double expansion coefficient $B_{nl}(k^2, \mu)$. 

The leading twist $2\pi$DAs is quoted as \cite{Diehl:1998dk,Diehl:2000uv,Polyakov:1998ze,Lehmann-Dronke:1999vvq} 
\beq &&\Phi_{\parallel,\perp}(u,\zeta,k^2) = 6u {\bar u} \sum_{n}^\infty \sum_{l}^{n+1} 
B_{nl}^{\parallel,\perp}(k^2, \mu) C_n^{3/2}(2u-1) C_l^{1/2}(2\zeta-1). \label{eq:2pidas-t2} \eeq 
Here $n$ is even (odd) and $l$ is odd (even) for the isovector (isoscalar) $\pi\pi$ state.  
The twist-three distribution amplitudes $\Phi^{(v,a)}_{\perp}, \Phi^{(s,t)}_{\parallel}$ are defined as the first time. 
For the isovector $\pi\pi$ state, they are parameterized in 
\beq &&\Phi^{(s)}_{\parallel}(u,\zeta,k^2) = 6u{\bar u} \Big[ B_{01}^\parallel(k^2, \mu) C_1^{1/2}(2\zeta-1) 
+ \frac{1}{6} \Big( B_{21}^\parallel(k^2, \mu) C_1^{1/2}(2\zeta-1) 
+ B_{23}^\parallel(k^2, \mu) C_3^{1/2}(2\zeta-1) \Big) C_2^{3/2}(2u-1) \Big], \nonumber\\
&&\Phi^{(t)}_{\parallel}(u,\zeta,k^2) = 3 \left( 2u-1\right)^2 B_{01}^\parallel(k^2, \mu) C_1^{1/2}(2\zeta-1)  \nonumber\\ 
&& \hspace{2.1cm} + \frac{3}{2} \left(2u-1\right)^2 \Big[ 5 \left(2u-1\right)^2 - 3 \Big] \Big[ B_{21}^\parallel(k^2, \mu) C_1^{1/2}(2\zeta-1)+ B_{23}^\parallel(k^2, \mu) C_3^{1/2}(2\zeta-1) \Big] C_2^{3/2}(2u-1), \nonumber\\
&&\Phi^{(v)}_{\perp}(u,\zeta,k^2) = 6u{\bar u} \Big[ B_{01}^\perp(k^2, \mu) C_1^{1/2}(2\zeta-1) 
+ \frac{1}{6} \Big( B_{21}^\perp(k^2, \mu) C_1^{1/2}(2\zeta-1)+ B_{23}^\perp(k^2, \mu) C_3^{1/2}(2\zeta-1) \Big) C_2^{3/2}(2u-1) \Big], \nonumber\\
&&\Phi^{(a)}_{\perp}(u,\zeta,k^2) = \frac{3}{4} \left[1 + \left( 2u-1\right)^2 \right] B_{01}^\perp(k^2, \mu) C_1^{1/2}(2\zeta-1) \nonumber\\
&& \hspace{2.1cm} + \frac{3}{7} \Big[ 3 \left(2u-1\right)^2 - 1 \Big] 
\Big[ B_{21}^\perp(k^2, \mu) C_1^{1/2}(2\zeta-1)+ B_{23}^\perp(k^2, \mu) C_3^{1/2}(2\zeta-1) \Big] C_2^{3/2}(2u-1). \label{eq:2pidas-t3} \eeq
To obtain the twist three isovector $2\pi$DAs, we have followed the conformal expansion of 
twist three distribution amplitudes of $\rho$ \cite{Ball:1998sk,Ball:1998ff,Ball:2007zt}. 
From the $C$-parity one can easily derive the symmetry property of isovector $2\pi$DAs that 
$\Phi^{I=1}(u, \zeta, k^2) = \Phi^{I=1}({\bar u}, \zeta, k^2)$. 
For the isoscalar $\pi\pi$ system, the double expansion of twist-three DAs reads as
\beq &&\Phi^{(s)}_{\parallel}(u,\zeta,k^2,\mu) = 6u{\bar u} \, C_1^{3/2}(2 u-1) 
\left[ B_{10}^\parallel(k^2, \mu) C_0^{1/2}(2\zeta-1) + B_{12}^\parallel(k^2, \mu) C_2^{1/2}(2\zeta-1) \right] \nonumber\\
&&\Phi^{(t)}_{\parallel}(u,\zeta,k^2,\mu) = C_1^{3/2}(2 u-1) \left[ B_{10}^\parallel(k^2, \mu) C_0^{1/2}(2\zeta-1) + B_{12}^\parallel(k^2, \mu) C_2^{1/2}(2\zeta-1) \right] .
\label{eq:2pidas-t3} \eeq
Here we consider only the asymptotic terms for the twist three isoscalar $2\pi$DAs. 
The $C$-parity indicates the symmetry property of isoscalar $2\pi$DAs that $\Phi^{I=0}(u, \zeta, k^2) = - \Phi^{I=0}({\bar u}, \zeta, k^2)$. 
At leading order in chiral perturbation theory (ChiPT), the scalar form factor is given by $\Gamma_\pi(s) = m^2 + {\cal O}(p^4)$ 
in the reliable region $\sqrt{k^2} \leqslant 0.1$ GeV \cite{Donoghue:1990xh}. 
Here, $m$ represents the chiral symmetry breaking term proportional to the quark masses, 
and the external momentum satisfies $p/\Lambda \sim p/(4 \pi f_\pi ) \leqslant 0.3$. 
For applications in heavy-flavor decays, the relevant kinematical region for $2\pi$DAs is the physical region where $\sqrt{k^2} > 2m_\pi$. 
The definitions in Eqs. (\ref{eq:2pidas-definition}) are therefore established in the chiral limit ($m_q \to 0$). 
This leads to a normalization condition for the scalar form factor: 
$ \langle \pi(k_1)\pi(k_2) \vert m_q {\bar q}(0) q(x) \vert 0 \rangle \stackrel{x \to 0}{\longrightarrow} \Gamma_\pi(k^2) \stackrel{m_q \to 0}{\longrightarrow} 0$.

\section{Expressions of specific partial wave form factors}

We collect the form factors at a certain partial wave as follow. 
\beq &&F_\parallel^{(l'=1,3,\cdots)}(q^2, k^2) = \frac{-i m_b^2 \sqrt{k^2}}{2 m_B^2 f_B} \sqrt{\frac{2l'+1}{2}} \frac{(l'-1)!}{(l'+1)!} 
\sum_{n=0,{\rm even}}^\infty \sum_{l=1,{\rm odd}}^{n+1} B_{nl}^\perp(k^2, \mu) J_n^\parallel(s_0,M^2,q^2,k^2)  I_{ll'}^\parallel, \label{eq:Fperp-1} \nonumber\\
&&F_\perp^{(l'=1,3,\cdots)}(q^2, k^2) = \frac{i m_b^2 \sqrt{\lambda k^2}}{2 m_B^2 f_B} \sqrt{\frac{2l'+1}{2}} \frac{(l'-1)!}{(l'+1)!} 
\sum_{n=0,{\rm even}}^\infty \sum_{l=1,{\rm odd}}^{n+1} B_{nl}^\perp(k^2, \mu) J_n^\perp(s_0,M^2,q^2,k^2) I_{ll'}^\parallel, \label{eq:Fperp-1} \nonumber\\
&& \sqrt{q^2} F_t^{(l'=1,3,\cdots)}(q^2, k^2) = \frac{m_b^2 }{2 m_B^2 f_B}\frac{\sqrt{2l'+1}}{2} \Big\{ 
\sum_{n=0,{\rm even}}^\infty \sum_{l=1,{\rm odd}}^{n+1} B_{nl}^\parallel(k^2, \mu) J_n^t(s_0,M^2,q^2,k^2) I_{ll'}^t \nonumber\\
&&\hspace{1cm} - \frac{m_b k^2}{f_{2\pi}^\perp} \int_{u_0}^1 du \left( \frac{1}{u^2} - \frac{m_b^2-q^2+u^2k^2}{u^3M^2} \right) 6u{\bar u}  
\left[ \left( B_{01}^\parallel C_0^{3/2}+ \frac{B_{21}^\parallel C_2^{3/2}}{6} \right) I^t_{1l'}
+ \frac{B_{23}^\parallel C_2^{3/2}}{6} I^t_{3l'} \right] e^{\frac{m_B^2-s(u,q^2)}{M^2}} \nonumber\\
&& \hspace{1cm} + \frac{m_b k^2}{f_{2\pi}^\perp} \frac{m_b^2-q^2+u_0^2k^2}{u_0 (s_0 - q^2)} 6 u_0 {\bar u}_0 
\left[ \left( B_{01}^\parallel C_0^{3/2}+ \frac{B_{21}^\parallel C_2^{3/2}}{6} \right) I^t_{1l'}
+ \frac{B_{23}^\parallel C_2^{3/2}}{6} I^t_{3l'} \right]_{u=u_0} e^{\frac{m_B^2-s_0}{M^2}} \Big\}, \label{eq:Ft-1} \nonumber\\
&&\sqrt{q^2} F_0^{(l'=1,3,\cdots)}(q^2, k^2) = \frac{\sqrt{\lambda}}{m_B^2-q^2-k^2} \sqrt{q^2} F_t^{(l')}(q^2, k^2) 
+ \frac{2 \sqrt{k^2} q^2  \beta_\pi }{m_B^2-q^2-k^2} \sum_{l =1}^\infty \sqrt{\frac{2l+1}{2}} F_\parallel^{(l=1,3,\cdots)}(q^2, k^2)I_{ll'}^0  \nonumber\\
&& \hspace{1cm} - \frac{i m_b k^2 q^2}{2 m_B^2 f_B f_{2\pi}^\perp} \frac{\sqrt{\lambda}}{m_B^2-q^2-k^2} 
\int_{u_0}^1 du \left( \frac{1}{u^2} - \frac{m_b^2-q^2+u^2k^2}{u^3M^2} \right) 
\left[ \Phi^{\parallel, ({\rm st})}_1(u) I^t_{1l'} + \Phi^{\parallel, ({\rm st})}_3(u) I^t_{3l'} \right] e^{\frac{m_B^2-s(u,q^2)}{M^2}} \nonumber\\
&&\hspace{1cm} + \frac{i m_b k^2 q^2}{2 m_B^2 f_B f_{2\pi}^\perp} \frac{\sqrt{\lambda}}{m_B^2-q^2-k^2} \frac{m_b^2-q^2+u_0^2k^2}{u_0 (s_0 - q^2)} 
\left[ \Phi^{\parallel, ({\rm st})}_1(u_0) I^t_{1l'} + \Phi^{\parallel, ({\rm st})}_3(u_0) I^t_{3l'} \right] e^{\frac{m_B^2-s_0}{M^2}},  \label{eq:F0-1} \nonumber\\
&&\sqrt{q^2} F_t^{(l'=0,2,\cdots)}(q^2, k^2) = \frac{m_b^2 }{2 m_B^2 f_B}\frac{\sqrt{2l'+1}}{2}\Big\{ 
\int_{u_0}^1 du \frac{m_b^2-q^2+u^2k^2}{u^2} \left[ B_{10}^\parallel I^t_{0l'} + B_{12}^\parallel I^t_{2l'} \right] 
6u {\bar u} C_1^{3/2} e^{\frac{m_B^2-s(u,q^2)}{M^2}} \nonumber\\
&&\hspace{1cm} - \frac{m_b k^2}{f_{2\pi}^\perp} \int_{u_0}^1 du \left( \frac{1}{u^2} - \frac{m_b^2-q^2+u^2k^2}{u^3M^2} \right) 
\left[ B_{10}^\parallel I^t_{0l'} + B_{12}^\parallel I^t_{2l'} \right] 6u {\bar u} C_1^{3/2} e^{\frac{m_B^2-s(u,q^2)}{M^2}} \nonumber\\
&& \hspace{1cm} + \frac{m_b k^2}{f_{2\pi}^\perp} \frac{m_b^2-q^2+u_0^2k^2}{u_0 (s_0 - q^2)} 
\left[ B_{10}^\parallel I^t_{0l'} + B_{12}^\parallel I^t_{2l'} \right] 6u_0 {\bar u}_0 C_1^{3/2}\vert_{u=u_0} e^{\frac{m_B^2-s(u,q^2)}{M^2}} \Big\}, 
\label{eq:Ft-2} \nonumber\\
&&\sqrt{q^2} F_0^{(l'=0,2,\cdots)}(q^2, k^2) = \frac{\sqrt{\lambda}}{m_B^2-q^2-k^2} \sqrt{q^2} F_t^{(l')}(q^2, k^2) 
+ \frac{2 \sqrt{k^2} q^2  \beta_\pi}{m_B^2-q^2-k^2} \sum_{l =1}^\infty \sqrt{\frac{2l+1}{2}} F_\parallel^{(l=2,\cdots)}(q^2, k^2)I_{ll'}^0  \nonumber\\
&& \hspace{1cm} - \frac{i m_b k^2 q^2}{2m_B^2 f_B f_{2\pi}^\perp} \frac{\sqrt{\lambda}}{m_B^2-q^2-k^2} 
\int_{u_0}^1 du \left( \frac{1}{u^2} - \frac{m_b^2-q^2+u^2k^2}{u^3M^2} \right) 
\Phi^{\parallel, ({\rm st})}(u) \left[ B_{10}^\parallel I^t_{0l'} + B_{12}^\parallel I^t_{2l'} \right] e^{\frac{m_B^2-s(u,q^2)}{M^2}} \nonumber\\
&&\hspace{1cm}+ \frac{i m_b k^2 q^2}{2m_B^2 f_B f_{2\pi}^\perp} \frac{\sqrt{\lambda}}{m_B^2-q^2-k^2} \frac{m_b^2-q^2+u_0^2k^2}{u_0 (s_0 - q^2)} 
\Phi^{\parallel, ({\rm st})}(u_0) \left[ B_{10}^\parallel I^t_{0l'} + B_{12}^\parallel I^t_{2l'} \right] e^{\frac{m_B^2-s_0}{M^2}}.  \label{eq:F0-2} \eeq
The integrations over the longitudinal momentum fraction $u$ are defined as  
\beq &&J_n^\parallel(s_0,M^2,q^2,k^2) = \int_{u_0}^1 du \frac{m_b^2 - q^2 + u^2k^2}{u^2 m_b f_{2\pi}^\perp} 6 u {\bar u} C_n^{3/2}(2u-1)  e^{\frac{m_B^2-s(u,q^2)}{M^2}}, \nonumber\\
&&J_n^\perp(s_0,M^2,q^2,k^2) = \int_{u_0}^1 du \frac{6 u {\bar u} C_n^{3/2}(2u-1)}{u m_b f_{2\pi}^\perp} e^{\frac{m_B^2-s(u,q^2)}{M^2}}, \nonumber\\
&&J_n^t(s_0,M^2,q^2,k^2) = \int_{u_0}^1 du \frac{m_b^2 - q^2 + u^2k^2}{u^2} 6 u {\bar u} C_n^{3/2}(2u-1) e^{\frac{m_B^2-s(u,q^2)}{M^2}}. \eeq
The integrations over the polar angle $\theta_\pi$ read as   
\beq
&&I_{ll'}^\parallel = \int_{-1}^1 d(\cos\theta_\pi) \frac{\sin \theta_\pi}{\beta_\pi \cos\theta_\pi} P_{l'}^{(1)}(\cos\theta_\pi) P_{l}^{(0)}(\beta_\pi \cos\theta_\pi),\nonumber \\
&&I_{ll'}^t= \int_{-1}^1 d(\cos\theta_\pi) P_{l'}^{(0)}(\cos\theta_\pi) P_{l}^{(0)}(\beta_\pi \cos\theta_\pi),\nonumber\\
&&I_{ll'}^0= \int_{-1}^1 d(\cos\theta_\pi) \frac{\cos\theta_\pi}{\sin \theta_\pi} P_{l'}^{(0)}(\cos\theta_\pi) P_{l}^{(1)}(\cos\theta_\pi). \eeq
In additional, the auxiliary distribution amplitudes are 
\beq
&&\Phi^{\parallel, ({\rm st})}_1(u) = \left( - 3 u {\bar u} + 6 u - 12 u^2 + 8 u^3 \right) B_{01}^\parallel \nonumber\\
&&\hspace{1cm} + \left[- u {\bar u} \left( -\frac{3}{4} + \frac{15}{4} (1-2u)^2 \right) 
+ 2 \left( 18 u - 171 u^2 + 774 u^3 - 1890 u^4 + 2556 u^5 - 1800 u^6 + \frac{3600}{7} u^7 \right) \right] B_{21}^\parallel \nonumber\\
&&\Phi^{\parallel, ({\rm st})}_3(u) = \left[ - u {\bar u} \left( -\frac{3}{4} + \frac{15}{4} (1-2u)^2 \right) 
+ 2 \left( 18 u - 171 u^2 + 774 u^3 - 1890 u^4 + 2556 u^5 - 1800 u^6 + \frac{3600}{7} u^7 \right) \right] B_{23}^\parallel \nonumber\\
&&\Phi^{\parallel, ({\rm st})}(u) = - 9 u {\bar u} (2u-1) + 6 u^2 - 6 u. \eeq

\section{Narrow width approximation and the $\rho$ resonance}

Our sum rules for the $B \to \pi\pi$ transition, derived from correlation functions Eq. (\ref{eq:cfs}) with $2\pi$DAs, 
do not rely on reproducing the $B \to \rho$ FFs in the kinematic limit as required in the $B$-meson LCSRs \cite{Cheng:2017smj}. 
This key difference stems from the choice of interpolating currents. 
Employing the current $j_b = i m_b \bar{b} \gamma_5 q_f$ isolates the $B$ meson and its excitations in the hadronic dispersion relation, in contrast to the electromagnetic current $j^\prime_\mu = \bar{d} \gamma_\mu u$ used in $B$-meson LCSRs, 
which selects the $\rho$ meson and its resonant partners. 

In the $B$-meson LCSRs, the derivation of $B \to \pi\pi$ FFs begins with a correlation function 
between the on-shell $B$-meson state and the vacuum, involving two bilocal currents, 
an electromagnetic current $j^\prime_\mu$ with momentum $k$, and a weak current $j_\nu$ with momentum $q$: 
\beq F_{\mu\nu} = i \int dx^4 e^{ik \cdot x} \langle 0 \vert {\bf T} \{ j^\prime_\mu(x), j_\nu(0)\} \vert {\bar B}^0(q+k) \rangle. \eeq
Considering, for example, the vector component of the weak current, 
the LCSR prediction for the $B \to \rho$ form factor $V$ is recovered from the $B \to \pi\pi$ form factor $F_\perp$ via a dispersion relation in $k^2$, 
retaining only the single $\rho$-pole contribution: 
\beq \frac{\sqrt{3} F_\perp^{(l=1)}(q^2,k^2)}{\sqrt{k^2} \sqrt{\lambda} } 
= \frac{g_{\rho\pi\pi} V(q^2) }{(m_B + m_\rho) \left[ m_\rho^2 - k^2- i \sqrt{k^2}\Gamma_\rho(k^2) \right] } + \cdots, \eeq
where the ellipsis denotes contributions from excited states. 
The strong coupling $g_{\rho\pi\pi}$ is defined via $\langle \pi^+(k_1) \pi^0(k_2) \vert \rho^+ \rangle = g_{\rho\pi\pi} \left( k_1 - k_2 \right)^\alpha \epsilon_\alpha$, 
and the $B \to \rho$ transition form factor is given by $\langle \rho(k) \vert {\bar u }\gamma_\nu b \vert {\bar B}(p) \rangle = \varepsilon_{\nu\alpha\beta\gamma} \epsilon^{\ast \alpha} q^\beta k^\gamma 2 V(q^2)/(m_B + m_\rho)$. 
Here, $\Gamma_\rho(k^2)$ is the energy-dependent total width of the $\rho$ meson.
Within the $\rho$-dominance approximation, the electromagnetic pion form factor 
entering the matrix element $\langle \pi^+(k_1) \pi^0(k_2) \vert  j^\prime_\mu \vert 0 \rangle = - \sqrt{2} \left( k_1 - k_2 \right) F_\pi(k^2)$ is taken as 
$F_\pi^\ast(k^2) = f_\rho g_{\rho\pi\pi} m_\rho/\sqrt{2}/( m_\rho^2 - k^2 + i \sqrt{k^2} \Gamma_\rho(k^2) )$, 
with the decay constant defining by $\langle \rho^+ \vert j^\prime_\mu \vert 0 \rangle = f_\rho m_\rho \epsilon^\ast_\mu$. 
In the zero-width limit ($\Gamma_\rho^{\rm tot} \to 0$), the sum rules of $P$-wave $B \to \pi\pi$ form factor $F_\perp^{(l=1)}(q^2,k^2)$ 
is deduced to the sum rules of $B \to \rho$ form factor $V$: 
\beq \frac{2 f_\rho m_\rho V(q^2)}{m_B + m_\rho} \int_{4m_\pi^2}^{s_0} ds e^{-s/M^2} 
\left(\frac{1}{\pi} \frac{\Gamma_\rho(k^2) \sqrt{k^2}}{(m_\rho^2 - k^2)^2 + s \Gamma_\rho^2(k^2)} \right) 
\longrightarrow \frac{2 f_\rho m_\rho V(q^2)}{m_B + m_\rho} e^{-m_\rho^2/M^2}. \eeq 

In the two-pion LCSRs calculation of $B \to \pi\pi$ transition, 
the $\rho$-meson DAs are incorporated into the $2\pi$DAs through a phase-shift analysis. 
Near the $\rho$ resonance, the $l=1$ phase shift is approximated by the Breit-Wigner formula 
\beq \delta_1(k^2 \sim m_\rho^2) = {\rm arctan}\left( \frac{m_\rho \Gamma_\rho}{m_\rho^2 - k^2} \right). \eeq
In the narrow-width limit, the resonant contribution to the phase space, given by the second term in Eq. (\ref{eq:Bnl}), produces the meson pole 
\beq \lim_{\Gamma_\rho \to 0} {\rm Exp} 
\left[ \frac{k^{2N}}{\pi} \int_{4m_\pi^2}^\infty ds \frac{\delta_1(k^2)}{s^N \left( s - k^2 - i 0 \right)} \right]
= {\rm Exp}\left[ i \frac{\pi}{2} - \ln\left(m_\rho^2- k^2 \right) + \ln m_\rho^2 + {\rm order} (N-1) \right] \eeq
The $\rho$-meson LCDAs are obtained by matching the double expansion of the $2\pi$DAs in Eq. (\ref{eq:2pidas-t2}) 
and the coefficient in Eq. (\ref{eq:Bnl}) near $k^2 \sim m_\rho^2$ to the Breit-Wigner form \cite{Polyakov:1998ze,Cheng:2019hpq}. 
For the leading-twist case, this yields  
\beq \phi_\rho = 6 u (1-u) \left[ 1 + \sum_{n = {\rm even}} a_n^\rho C_n^{3/2}(2u - 1) \right]. \eeq 
The Gegenbauer coefficients $a_n^\rho$ are expressed in terms of the double expansion coefficients $B_{n1}(k^2)$ at low energies: 
\beq a_n^\rho = B_{n1}(0) \, {\rm Exp} \left[ \sum_{m=1}^{N-1} \, \frac{1}{k!} \frac{d^m}{dk^{2m}} 
\Big[ \ln B_{nl}(k^2) - \ln B_{l-1 l}(k^2) \Big] \Big\vert_{k^2 \to 0} m_\rho^{2m} \right], \eeq 
and the decay constant is given by $f_\rho = \sqrt{2} \Gamma_\rho \mathrm{Im} B_{01}(m_\rho^2)/g_{\rho\pi\pi}$. 
Although we have shown explicit results for the leading-twist DA, the phase-shift analysis applies universally, 
so higher-twist $\rho$-meson LCDAs can be derived in a similar manner. 


\end{document}